# CTOF measurements and Monte Carlo analyses of neutron spectra for the backward direction from a lead target irradiated with 200 to 1000 MeV protons

I. L. Azhgirey, V. I. Belyakov-Bodin, I. I. Degtyarev

*Institute for High Energy Physics, Protvino, Russian Federation*

V. A. Sherstnev

*Institute of Theoretical and Experimental Physics, Moscow, Russian Federation (former affiliation)*

S. G. Mashnik

*Los Alamos National Laboratory, Los Alamos, NM 87545, USA*

F. X. Gallmeier and W. Lu

*Oak Ridge National Laboratory, Oak Ridge, TN 37830, USA*

## Abstract

A calorimetric-time-of-flight (CTOF) technique was used for real-time, high-precision measurement of neutron spectrum at an angle of $175^o$ from the initial proton beam direction, which hits a face plane of a cylindrical lead target of 20 cm in diameter and 25 cm thick. A comparison was performed between the neutron spectra predicted by the MARS, RTS&T, MCNP6, and the MCNPX 2.6.0 transport codes and that measured for 200, 400, 600, 800, and 1000 MeV protons.

**1. Introduction**
The MARS, RTS&T, MCNP6, and the MCNPX 2.6.0 code systems are useful tools for the analysis and design of systems incorporating high-intensity neutron sources such as an accelerator-driven system (ADS). These codes simulate the transport of particles in matter with cascading secondary particles over a wide energy range. This encompasses the ability to determine the neutron spectrum around a system. For many system designs, and especially for high-energy, high beam current applications, an important design factor is the neutron spectrum emitted from the ADS target. Consequently, it is important that the accuracy for neutron spectrum predictions by the codes is well determined. The goal of this study is to determine the precision of the MARS, RTS&T, MCNP6, and the MCNPX 2.6.0 transport codes in making these predictions at a wide angle.

The geometry for all codes which modeled the experimental configuration consists of a cylindrical target and a detecting volume. The aim of our work was to investigate neutron spectra only in backward direction; therefore we chose 175° as a suitable angle. There are several reasons why we limited ourselves to only the backward angle of 175°. First, we need such information for programmatic needs, for shielding considerations, to be able to prevent cases when personnel may receive radiation from backward fluxes, as once happened at BNL. Second, it is much more difficult for all models to describe particle production at very backward angles than at intermediate or forward angles. Our data allow us to test event-generators used by transport codes in this "difficult" kinematics region. Third, spectra of secondary particles at very backward angles are of great academic interest, to understand the mechanism of cumulative particle production, under investigation for almost four decades, but still with many open questions.

## 2. Facility description and operation principle

A detailed description of the experimental facility construction, the calorimetric-time-of-flight (CTOF) technique, method of data taking, and experimental results for tungsten and iron targets irradiated by protons was recently given [1-3]. Therefore, only a brief description of the ZOMBE facility and the CTOF technique are needed for understanding the problem is presented in this paper.

We used a proton beam extracted from the U-1.5 ring accelerator (booster) in the Institute for High Energy Physics, Protvino, Russian Federation, by a tripping magnet. The booster cycle time was about 9 s containing 29 pulses of about 30 ns long, each containing $1.3 \times 10^{11}$ protons. Extraction time is 1.6 s. Beam intensity measurements were carried out for each pulse by using an induction current sensor [4] with an accuracy of about 5%. The proton beam energy values were determined with an error of approximately 0.5% in the proton energy value. The average radial distribution of protons in the beam has been obtained for this experiment [1-3] and was compatible with a Gaussian of a full-width at half-maximum of 24 mm. A proton beam hits a face plane of target placed on the ZOMBE facility (see Fig. 1). The target axis position was superimposed on the geodesic beam axis.

The CTOF technique consists of a set of current mode time-of-flight instruments each having a set of organic scintillation detectors placed at a different distance from a source at an angle of interest away from the proton beam direction. One - litre (10 x 10 x 10 $cm^3$) organic scintillators were used as the detectors. The first detector was placed at distance 8.7 m from the face plane of the target at 175º to the proton beam direction. The second detector was placed at distance 29 m at the same angle. There are only two directions into the booster-room (175º and 96º) where long-distance (more then 26 m) detectors may by placed for measurements of high-energy neutrons. Wavelength shifters were coupled with the scintillators and its output signals were transported to each photo-multiplier tube (PMT) by a 120 m long light-guide (LG). The PMT outputs were digitized by a TDS 3034b oscilloscope (0.4 ns resolution, 200 MHz bandwidth). The data were obtained within a time of 2000 ns from the start of the irradiation cycle and sent to an off-line computer for further analyses. Since the oscilloscope resolution is 0.4 ns, there are so many experimental points that the measured neutron spectra in Figs. 3-12 appear as continuous lines.

All of the beam pickup signals were digitized for reconstruction of the average proton beam intensity as a function of time. In the measurements, 5000 proton beam pulses were used to average the photometric signals. To exclude any radiation-induced noise from the photometric signals of the detectors, such as neutrons reflected from the walls, ceiling and floor, radiations from the accelerator, and so on, it was necessary to make additional measurement for the case when the burst source overshadowed for the detectors cylindrical shield. For the shield we used a set of 45 cm thick iron and 40 cm thick tungsten discs, both 20 cm in diameter. We have the option of using only one shield in some experiments, so we used one for the 175º-direction in this experiment.

The difference of these two measurements will give a pure signal from the source. The amplitude of a light signal in a detector caused by a nonzero-mass particle is:

$$u(t) = k \cdot S(E) \cdot e_f(E) \cdot dE/dt \, ,$$

where $S(E)$ is the energy spectrum of particles emitted from a target, $e_f(E)$ is the sensitivity of the detector, *dE/dt* is the time derivative of the particle energy for *r*-distance detector, and *k* is a coefficient. The sensitivity $e_f(E)$ of the detector to neutrons (and cosmic μ-meson ($e_\mu$)) was calculated by the RTS&T code [4] and is presented in previous our articles [1-3]. We used the time integral of a photometric signal of the scintillator caused by a cosmic μ-meson signal for determination of the *k*-coefficient, i.e. for absolute calibration of the detectors.

The digitized voltage amplitude is determined by the following integral equation:

$$U(t) = u(t)^* \cdot p(t)^* \cdot d(t)^* \cdot LG(t)^* \cdot PMT(t), \qquad (1)$$

where *p(t)* is the proton beam intensity as a function of time; *d(t), LG(t)*, and *PMT(t)* are pulse functions of the detector, LG, and PMT, respectively; and the asterisk (*) is a convolution transform symbol:

$$u(t)^* \cdot p(t) = \int_o^t u(\xi) \cdot p(t - \xi)\, d\xi.$$

An original code was made for determining *u(t)* function from the Eq. (1) by using *U(t), p(t), d(t), LG(t),* and *PMT(t)*.

The part of photometric signal caused by gamma rays from the target we determined from the distant detector signal, which could be separated easily and clearly into gamma rays and nucleon components. Moreover, for reconstruction of neutron spectra in this experiment we used data only for the distant detector since analysis experimental data and called on preliminary calculation has shown negligible contribution of the proton-induced signal in the general photometric signal from this detector. However, this part of our study is outside the aim of this work and will be published in a future paper.

The total relative error including beam intensity, beam energy, detector sensitivity, and photometric signal errors for 400 and 1000 MeV protons are presented on Fig.2.

## 3. Monte Carlo simulation

The MCNPX calculated backscattering neutron spectra in Figs. 3, 5, 7, 9 and 11 were performed with version 2.6.0 [6] using the Bertini INC [7] coupled with the Dresner evaporation model [8] (MCNPX defaults, using also the multistage pre-equilibrium model, MPM [9]), the Liege INC model INCL4 [10] merged with the GSI evaporation/fission model

ABLA [11], as well as the 03.01 version of the Cascade-Exciton Model event-generator CEM03.01 [12, 13] using an improvement of the Generalized Evaporation Model GEM2 by Furihata [14] to calculate evaporation/fission and its own Modified Exciton Model (MEM) [15] in its latest version as described in Refs. [12,13]. For the proton and neutron transport and interactions up to 150 MeV, MCNPX in all cases uses applied tabulated continuous-energy cross sections from the LA150 proton and neutron libraries. The calculation assumed a Gaussian-distributed incident proton beam with a FWHM of 2.4 cm hitting a cylindrical iron target. The radius of the lead target is 10 cm and the length is 25 cm. The backscattered neutrons were tallied at an angle of 175$^o$ of the incident beam direction and at 6 m upstream of the front surface of the target. A ring surface tally was adopted to improve the efficiency of the simulation by taking advantage of the symmetry.

MCNP6 [16] is the latest and most advanced LANL transport code representing a merger of MCNP5 [17] and the latest version of MCNPX [18]. MCNP6 is still under development with a plan to make it available to users via RSICC at ORNL, Oak Ridge and NEA/OECD in Paris in 2011. The version of MCNP6 used here has implemented the latest modification of the Cascade-Exciton Model event generator CEM03.02 [13, 19] as well as the Los Alamos version of the Quark-Gluon String Model code LAGGSM03.01 [13, 20], used mostly as a very high-energy event generator and to describe reactions induced by heavy-ions. Here, we test and validate MCNP6 with both CEM03.02 and LAQGSM03.01 event generators, using the latest versions of the data libraries we have currently at LANL to describe transport and interaction of nucleons at energies below 150 MeV. The geometry of the problem, the source (proton beam), and the tally (results) simulated by MCNP6 are exactly the same as used by MCNPX and described above.

The MARS code [21] simulates a process of development of the nuclear-electromagnetic cascades in matter. Its physical module is based mostly on parameterization of the physical processes. It provides flexibility and rather high operation speed for engineering applications and optimization tasks. MARS is used for radiation-related modeling at accelerators, such as shielding design [22], dose distribution and energy deposition simulations [23], and radiation background calculations [24]. In this region of the primary proton energy (below 5000 MeV), MARS uses a phenomenological model for the production of secondary particles in inelastic hadron-nucleus interactions [25]. For low-energy neutron transport at energies below a 14.5 MeV threshold and down to the thermal energy, MARS uses a multi-group approximation and a 28 group library of neutron constants BNAB-MICRO [26].

In the MARS cascade model, charged particles (protons and pions) are absorbed locally when their kinetic energy falls below 10 MeV. At an energy of 10 MeV, proton and pion ionization ranges in lead are more than two orders of magnitude less than the average path before inelastic interaction with the nuclei. Therefore for our initial beam energies from 200 to 1000 MeV we neglect secondary interactions caused by sub-threshold charged hadrons, because it can contribute only small additions to the neutron spectrum, formed mainly by the interactions of high-energy particles.

The target was considered as a solid lead cylinder with a length of 25 cm and diameter of 20 cm. A detecting volume of 10x10x10 cm$^3$ was placed 9 m upstream of the target front face plane, and the angle between beam direction and direction from target to detector is 175°. A Gaussian distribution of the proton beam with $\sigma_{x,y}$ = 1.4 cm was used, which is an approximation of the measured distribution. Test runs show no difference in spectra between a point-like beam and a realistic distribution. The space between the target and detector was filled with air.

In the RTS&T calculations, the hadron-induced nuclear reaction process in the energy region about 20 MeV to 5 GeV is assumed to be a three-step process of spallation (intranuclear cascade stage), pre-equilibrium decay of residual nucleus and the compound nucleus decay process (evaporation/high-energy fission competition). To calculate the intra-nuclear cascade stage, the Dubna version of intra-nuclear cascade model [27] coupled with the Lindenbaum-Sternheimer isobar model for single- and double-pion production in nucleon-nucleon collisions and single-pion production in pion-nucleon collisions was provided. Recently, an addition of multiple-pion channels was included in code package to simulate up to 5 pions emission. The pre-equilibrium stage of nuclear reaction simulation is based on the exciton model. As proposed in [28], the initial exciton configuration for pre-equilibrium decay is calculated at the cascade stage of reaction or postulated in the general input. The equilibrium stage of reaction (evaporation/fission processes competition) is performed according to the Weisskopf-Ewing statistical theory of particle emission and Bohr and Wheeler or Fong theories of fission. To calculate the quantities determining the total fission width, Atchison prescriptions are used. The RTS&T code uses continuous-energy nuclear and atomic evaluated data files to simulate of radiation transport and discrete interactions of the particles in the energy range from thermal energy up to 20/150/3000 MeV. In contrast with the MCNP, the ENDF-data driven model of the RTS&T code does access the evaluated data directly. In current model development, all data types provided by ENDF-6 format can be used in the coupled many-particle radiation transport modeling. Universal data reading and preparation procedure allows us to use various data library written in the ENDF-6 format (ENDF/B, JENDL, JENDL-HE, FENDL, CENDL, JEF, BROND, LA150, ENDF-HE/VI, IAEA Photonuclear Data Library etc.). ENDF data pre-processing (linearization, restoration of the resolved resonances, temperature dependent Doppler broadening of the cross sections and checking and correcting of angular distributions and Legendre coefficients for negative values are produced automatically with the Cullen's ENDF/B Pre-processing codes [29] LINEAR, RECENT, SIGMA1 and LEGEND rewritten in ANSI standard FORTRAN-90. ENDF-recommended interpolation laws are used to minimize the amount of data. For data storage in memory and their further use, a dynamically allocated tree of objects is organized. All types of reactions provided by ENDF-6 format are taken into account for the particle transport modeling. More details on the RTS&T code may be found in [5] and references therein.

## 4. Comparison of experimental and calculation results

As shown in Figs. 4, 6, 8, 10 and 12, MARS results agree quite well with the measured spectra and the MCNP6 results. MARS overestimates a little the 20-80 MeV portion of the spectra at 200 MeV (and to a less degree, also at 400 MeV) and underestimates the high-energy tails of the spectra at 600, 800, and especially at 1000 MeV. All event-generators of MCNPX underestimate a little the measured spectra in the ~ 5-10 MeV energy region, as well as at high energies, at the very end of the spectra tails, and also MCNPX underestimates a little the measured spectra at very low energies, in the 1-3 MeV region. The overall agreement of results by different event-generators with the measured spectra depends mostly on the proton incident energy.

A small difference is observed for neutron energies from several MeV to ~50 MeV, where a shoulder is present on the measured spectra around 5 MeV showing higher backscattered neuron flux compared to the MCNPX calculated spectra for neutrons from several MeV to 10 MeV and lower neutron flux for neutrons above 10 MeV up to ~50 MeV. Neither event-generators (CEM03.01 and Bertini+Dresner) using their own, different pre-equilibrium models, nor the INCL+ABLA, which does not use a pre-equilibrium stage, are able to capture this shoulder. Such a shoulder on the measured spectrum probably indicates the superposition of the flux from the low-energy evaporation neutrons and high-energy cascade neutrons. We can observe that all models of MCNPX agrees quite well with the measured neutron spectra from lead at neutron energies above several MeV, for all incident proton energies studied here. But all the MCNPX calculations underestimate by up to a factor of about 1.5 the low portion (below ~ 2 MeV) of the measured neutron spectra, at all incident proton energies, just like provided also by the RTS&T code. We do not see such an underestimation for the results by MCNP6 (which agree better with the data) neither for the Bertini+Dresner and INCL+ABLA event generators, exactly the same as the ones used in MCNPX, nor for CEM03.02 of MCNP6, which is an improvement over the CEM03.01 version used by MCNPX 2.6.0. In the case of Bertini+Dresner and INCL+ABLA event generators, the only significant difference between the MCNPX and MCNP6 calculations is the use of different data libraries, which should affect mostly just the low energy portions of the calculated spectra. From here, we may assume that the observed underestimation by MCNPX of the low-energy portions of the measured spectra and the small disagreement with the MCNP6 results for neutron energies below ~2 MeV are related to the different data libraries used by MCNPX and MCNP6, though this question is quite serious and requires a further investigation of the transport of particles in the matter as described by the two codes, outside the aim of the current work.

MARS underestimates the high-energy tails of spectra at the highest beam energies. It produces approximately three times less neutrons compared to the data. Such behavior for this kinematical region is similar to that for a tungsten [2] and an iron [3] targets.

We now comment on a still open question. Our data indicate a small shoulder around 4 to 10 MeV which is seen in practically all measured neutron spectra. A similar situation was observed also for a tungsten target [2] and also for a iron target [3], and none of the models we tested so far reproduced well this feature. The current calculations for lead by MCNP6 tend to agree better with the measured neutron spectra in this "shoulder" region than results by other codes and than the previous calculations for the tungsten [2] and iron [3] targets, though even MCNP6 does not reproduces completely well this feature.

We do not have yet a good understanding of this situation, as several combined effects could contribute to this feature for thick targets. Further investigations, including using other nuclear reaction models and probably more measurements are needed to solve this problem. We cannot solve it by comparing our results with previous measurements and calculations simply because we do not know of any previous studies of neutron spectra from thick targets at 175º. To the best of our knowledge, only the SATURNE measurements by David et al. [30] on thick lead targets at 800, 1200, and 1600 MeV analyzed with the INCL4+ABLA and Bertini+Dresner event-generators are similar to our work. The largest angle measured at SATURNE was only 160º, and the target dimensions were different, so we cannot compare directly our results with Ref. [30]. Similar measurements of neutron spectra from a thick lead target bombarded with 0.5 and 1.5 GeV protons ware done by Miego et al. at KEK [31], but the largest angle measured at KEK was of only 150º, and the target dimensions were different, so we cannot compare directly our results with Ref. [31] as well. Neutron spectra from thin lead targets bombarded by protons of similar to our energies were measured by about a dozen of different groups. Recent compilations of references to these measurements are tabulated by Trebukhovsky et al. [32] and by Yurevich [33] and many useful details may be found in the recent very comprehensive spallation handbook by Filges and Glodenbaum [34]. But the largest angle measured in all known to us thin-target experiments was of only 160º, so we cannot compare our data even with the measured thin-target neutron spectra.

## Acknowledgements

We are grateful to G.A. Losev and A.V. Feofilov (both of the Institute for High Energy Physics) for providing the beam so reliably under the conditions demanded for this experiment. This work was partially supported by the US DOE.

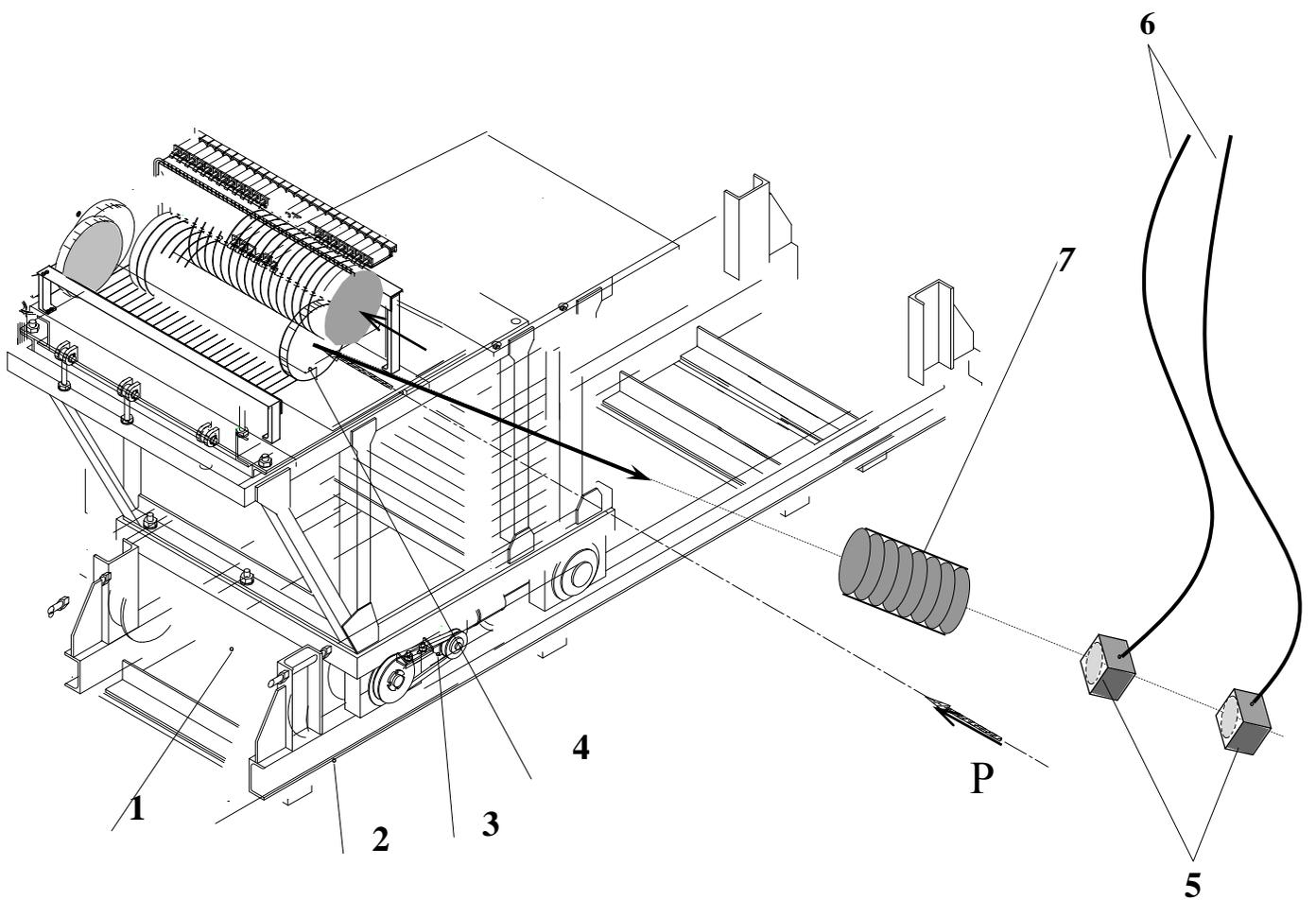

**Fig. 1.** Scheme of the ZOMBE facility: (1) mobile test bench; (2) rails; (3) bench displacement motor; (4) target; (5) detectors; (6) light-guide; (7) local shield; and $P$ - proton beam direction.

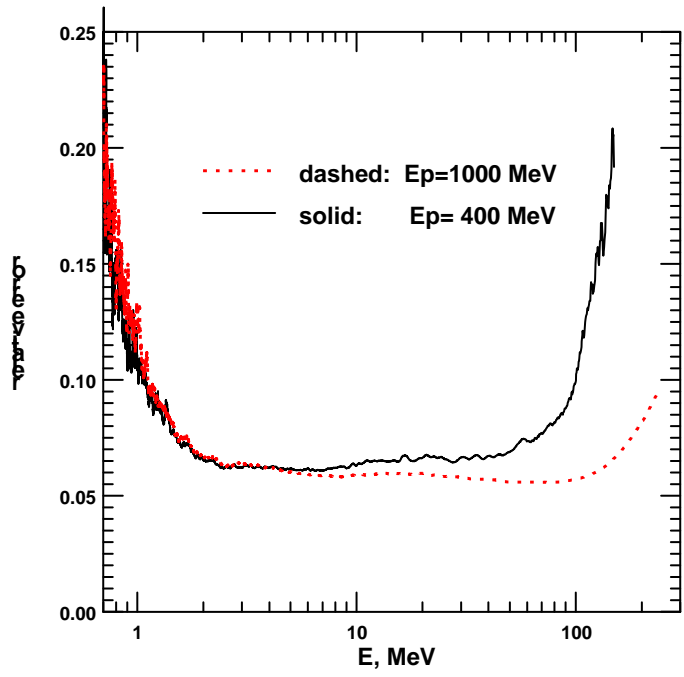

**Fig. 2.** The total relative error.

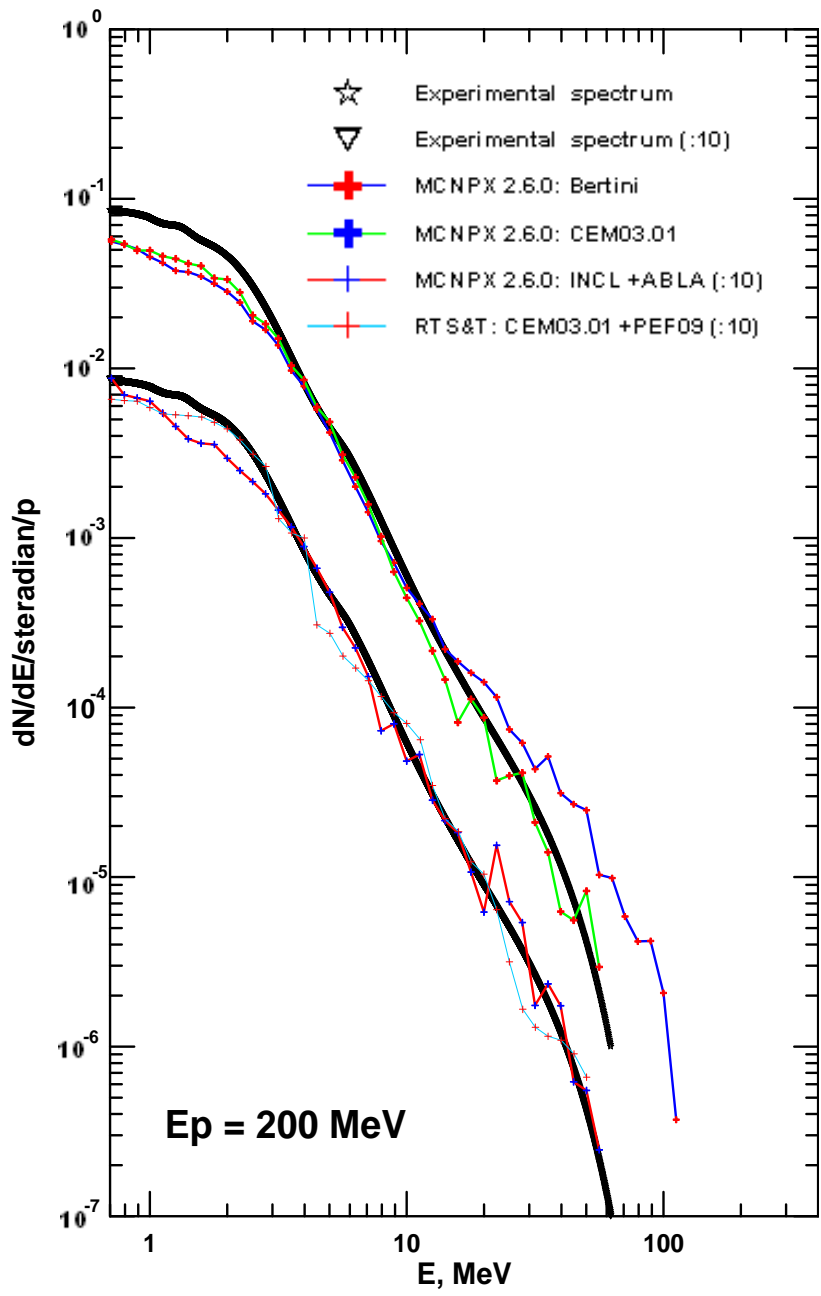

**Fig. 3.** Experimental neutron spectrum for 200 MeV initial proton energy and calculations by the RTS&T and the MCNPX 2.6.0 codes.

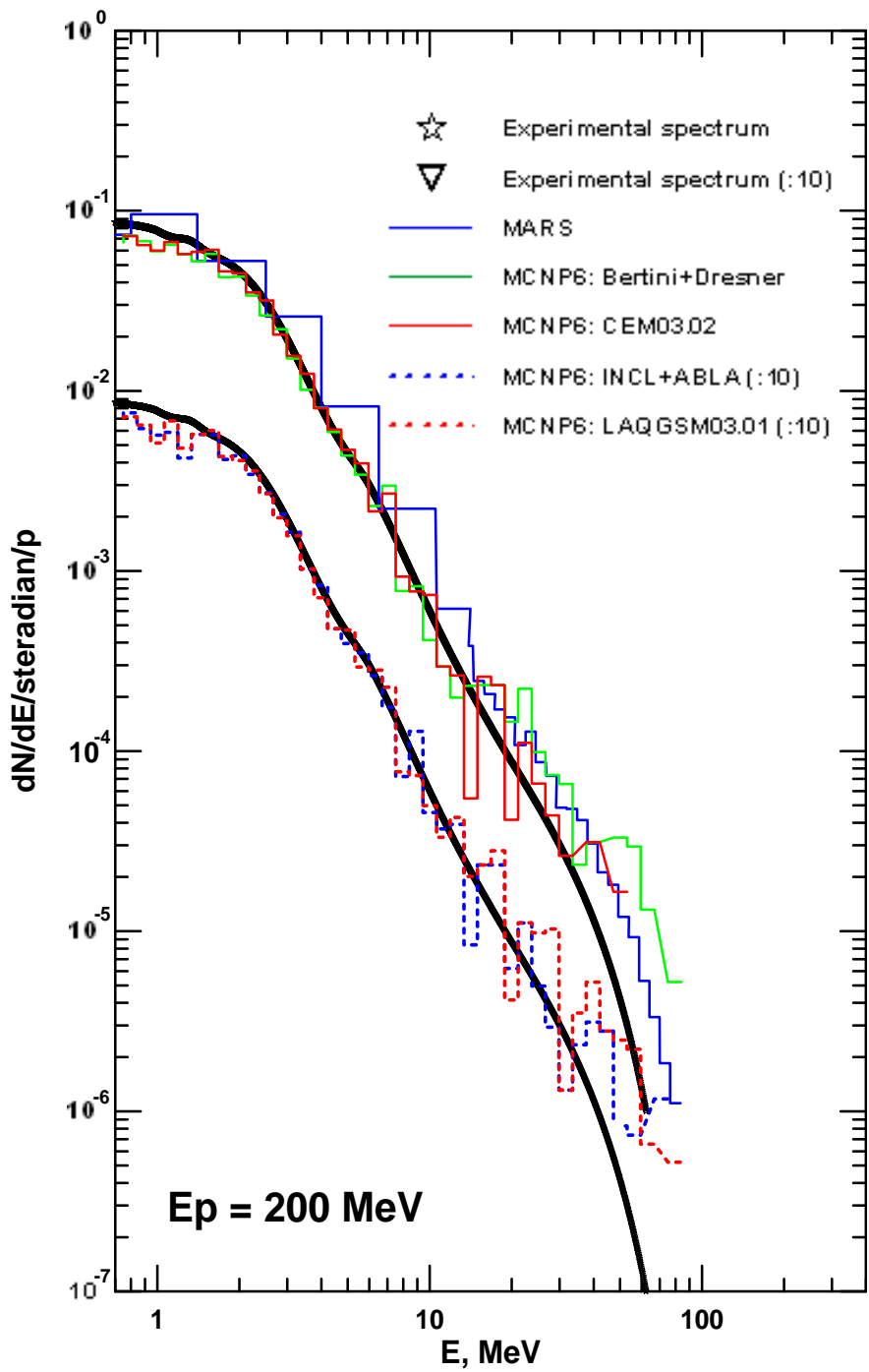

**Fig. 4.** Experimental neutron spectrum for 200 MeV initial proton energy and calculations by the MARS and the MCNP6 codes.

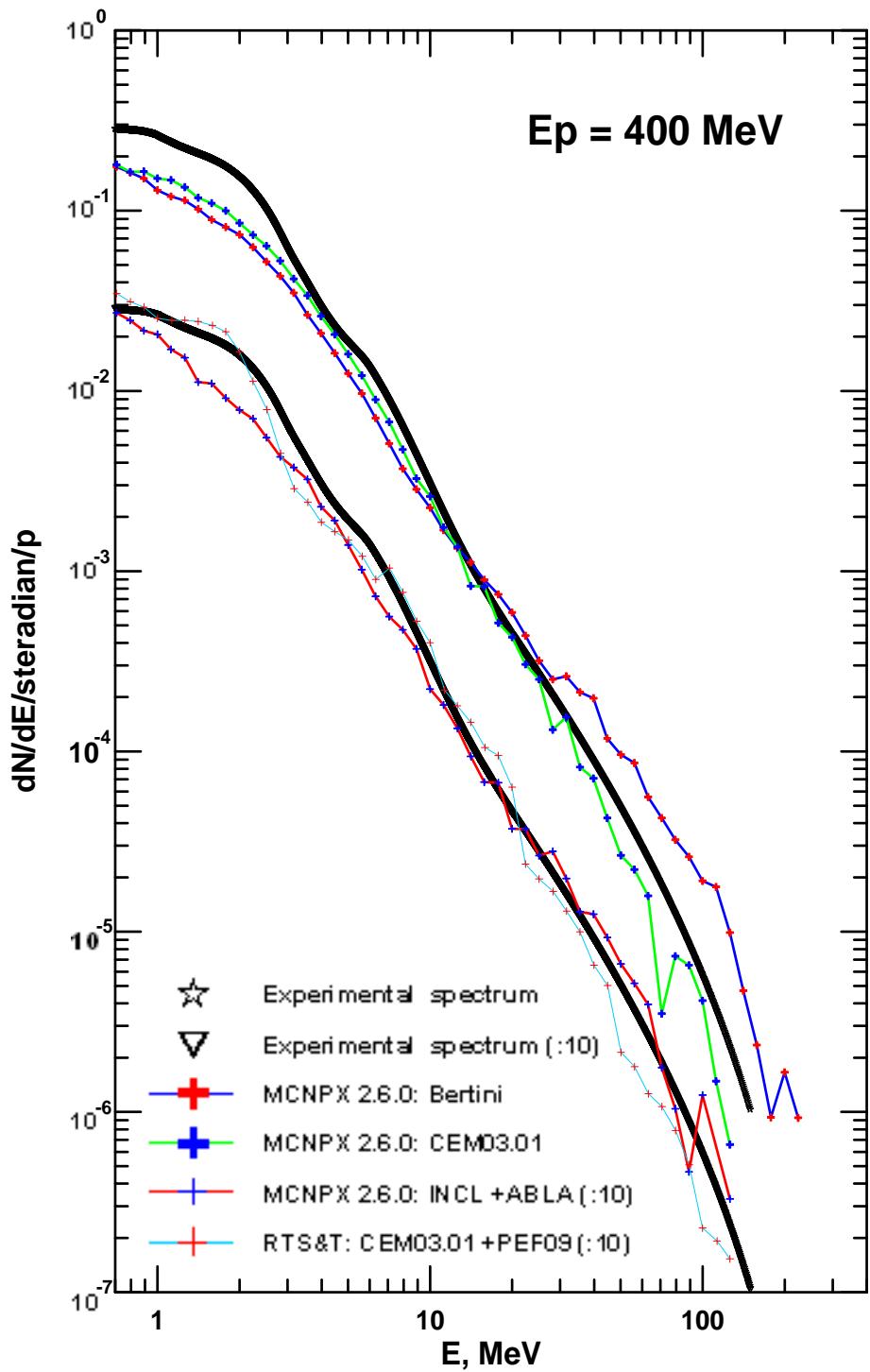

**Fig. 5.** Experimental neutron spectrum for 400 MeV initial proton energy and calculations by the RTS&T and the MCNPX 2.6.0 codes.

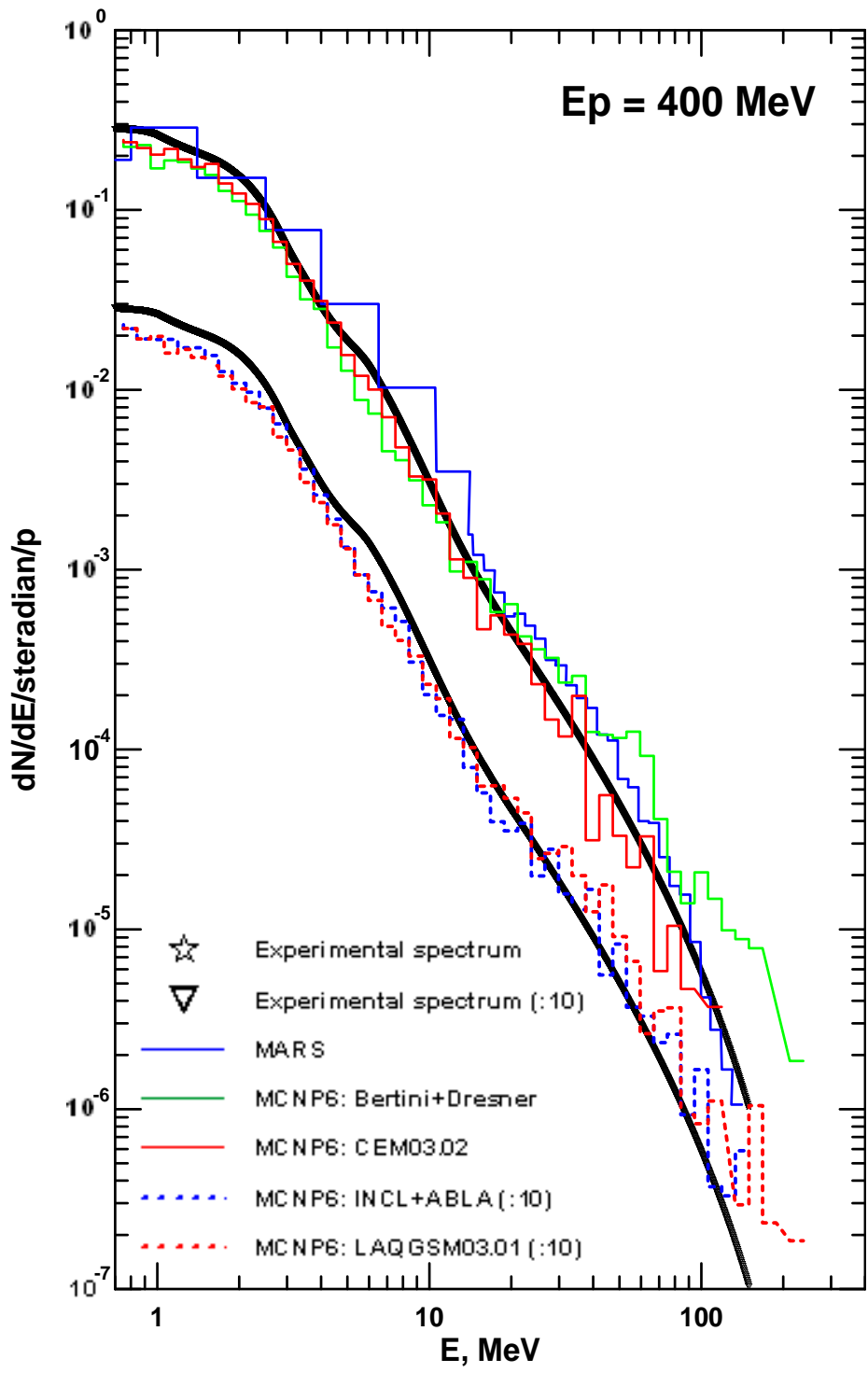

**Fig. 6.** Experimental neutron spectrum for 400 MeV initial proton energy and calculations by the MARS and the MCNP6 codes.

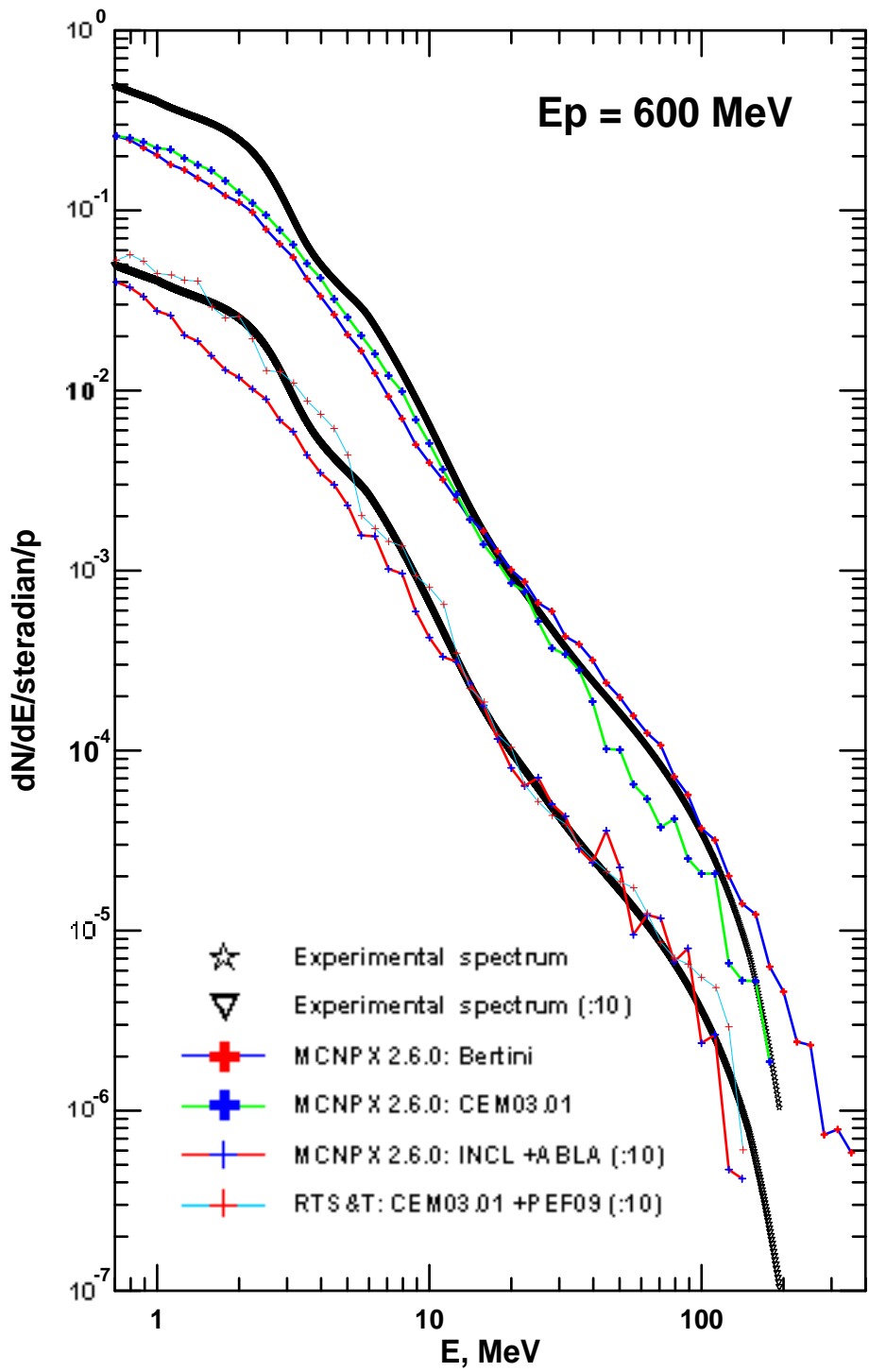

**Fig. 7**. Experimental neutron spectrum for 600 MeV initial proton energy and calculations by the RTS&T and the MCNPX 2.6.0 codes.

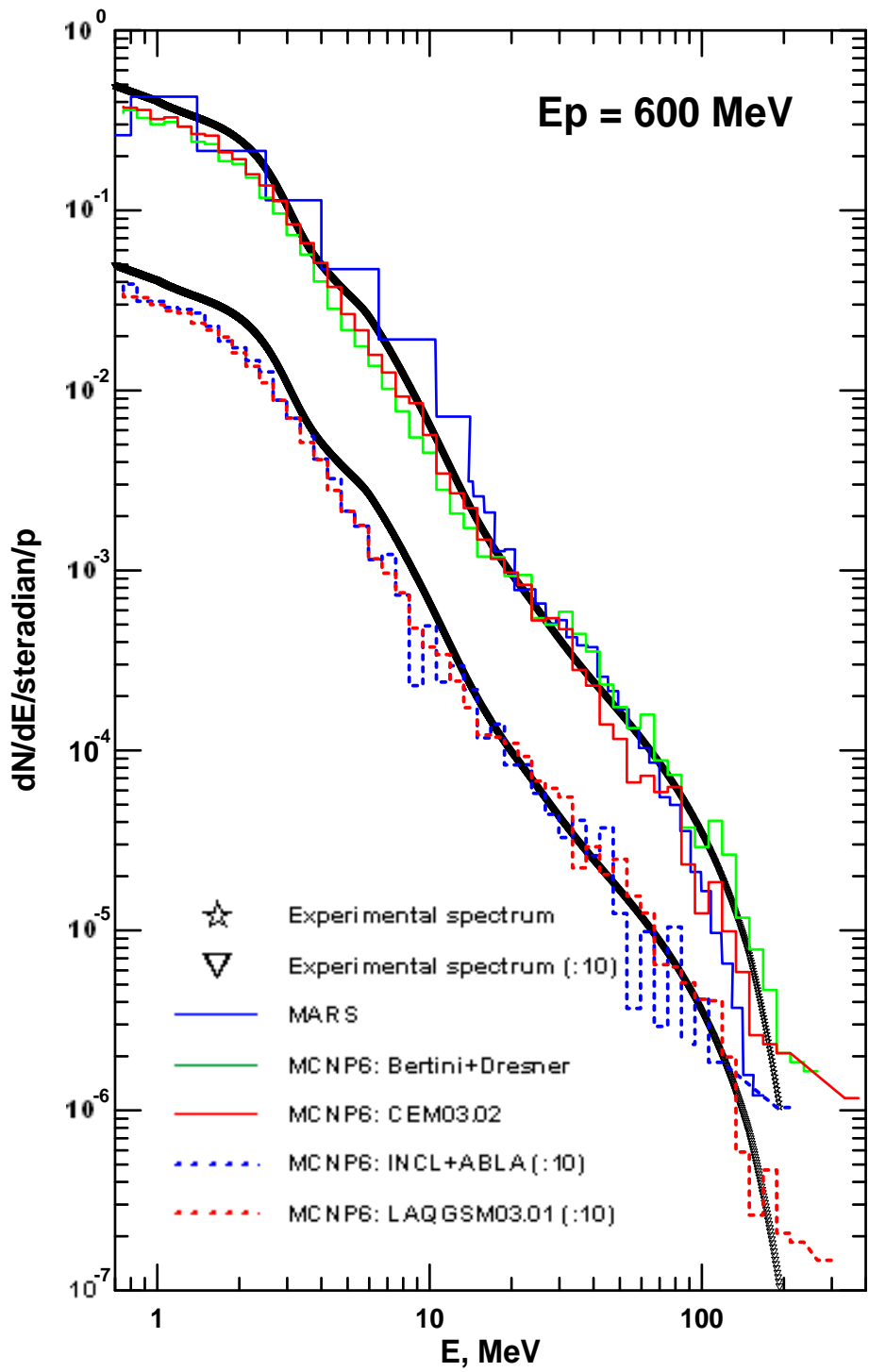

**Fig. 8.** Experimental neutron spectrum for 600 MeV initial proton energy and calculations by the MARS and the MCNP6 codes.

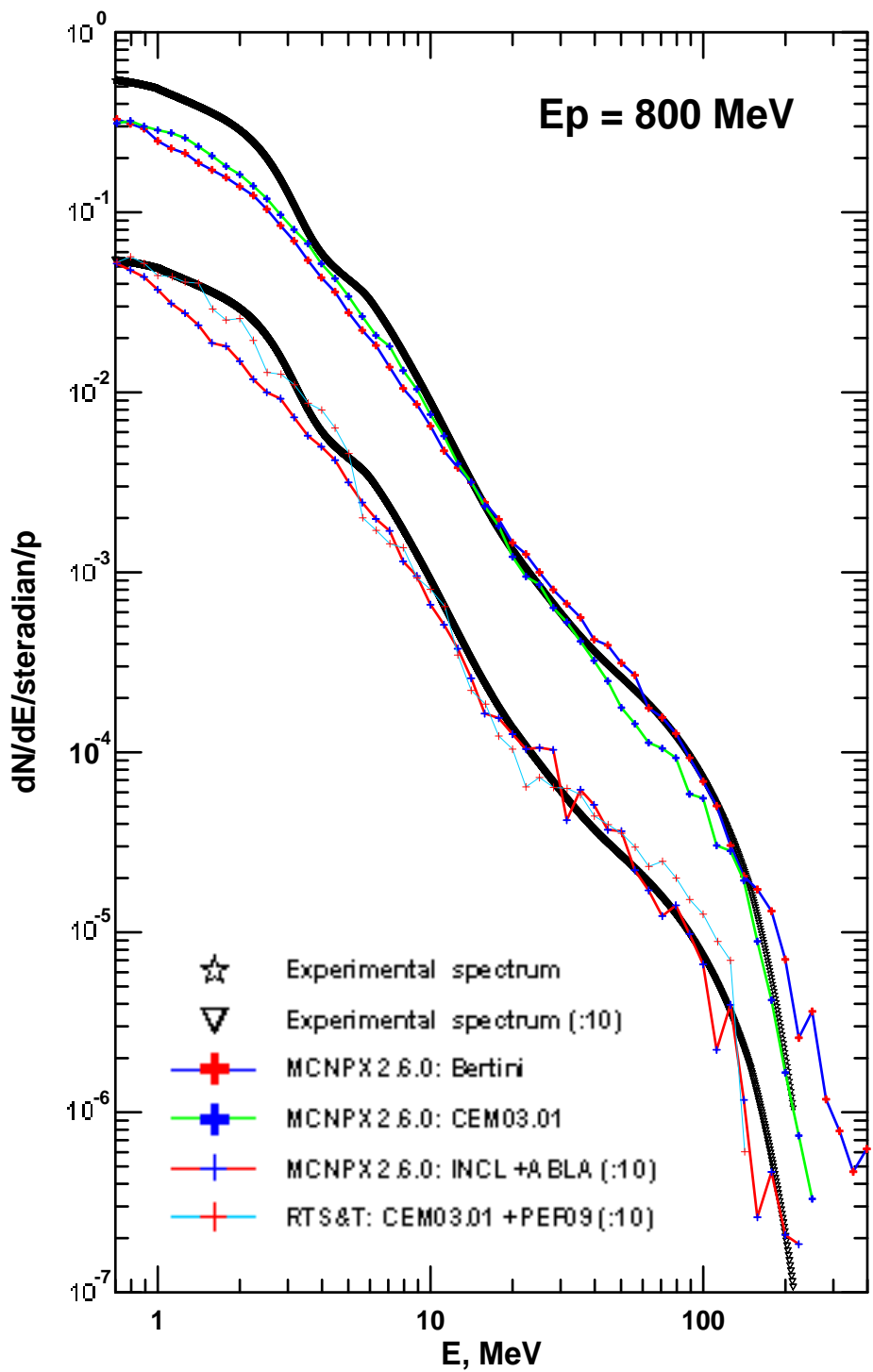

**Fig. 9.** Experimental neutron spectrum for 800 MeV initial proton energy and calculations by the RTS&T and the MCNPX 2.6.0 codes.

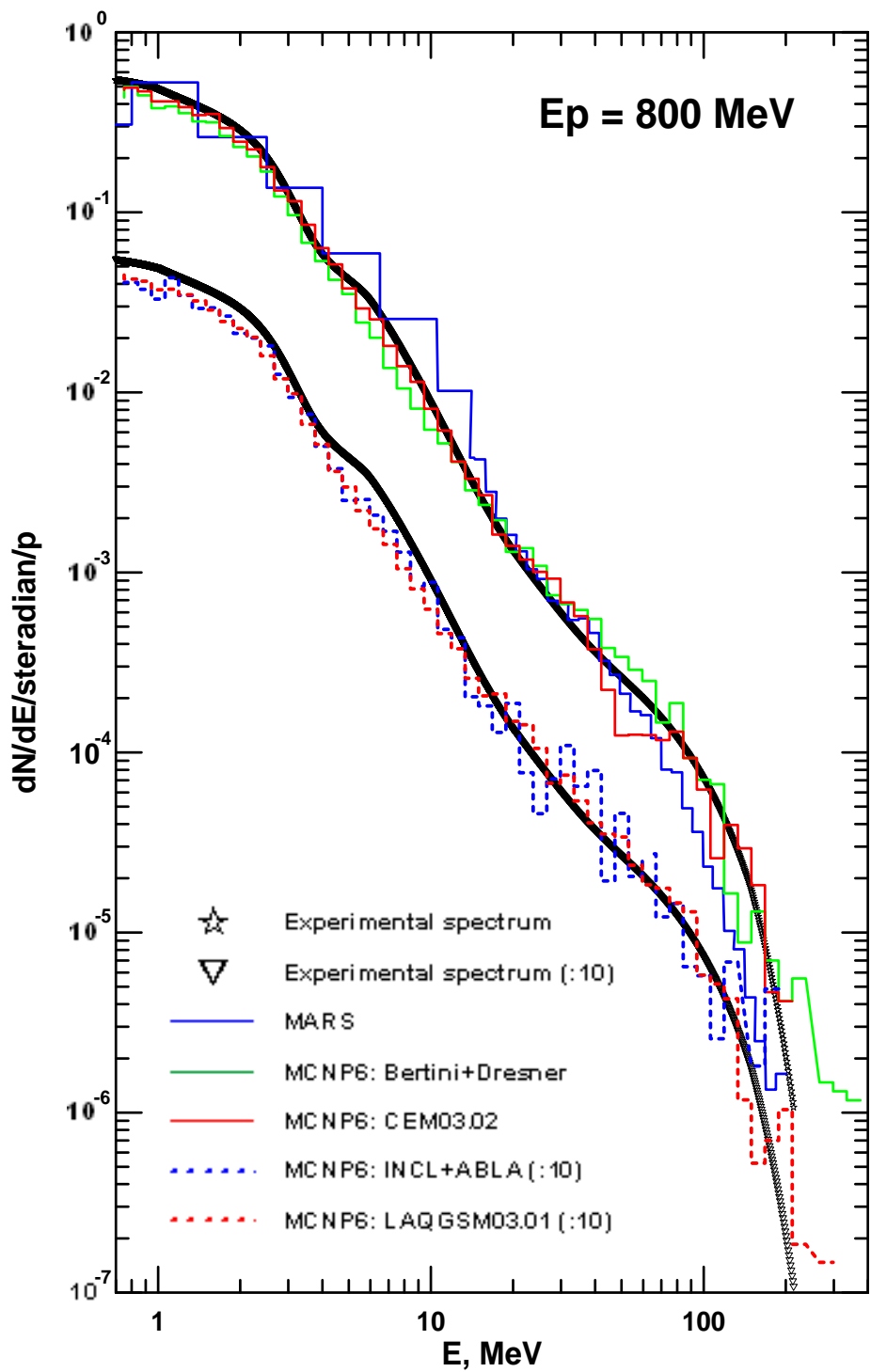

**Fig. 10.** Experimental neutron spectrum for 800 MeV initial proton energy and calculations by the MARS and the MCNP6 2.6.0 codes.

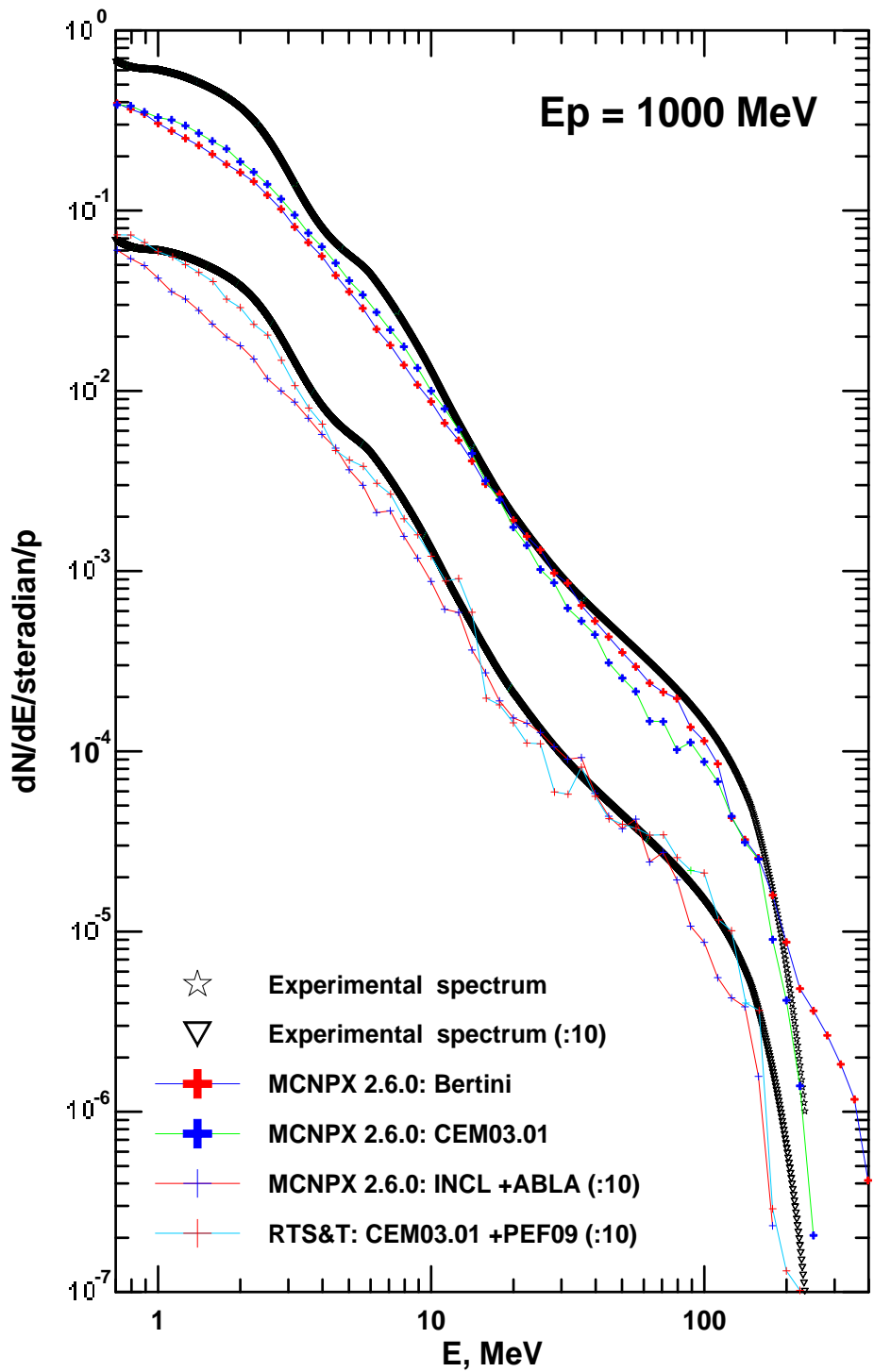

**Fig. 11.** Experimental neutron spectrum for 1000 MeV initial proton energy and calculations by the RTS&T and the MCNPX 2.6.0 codes.

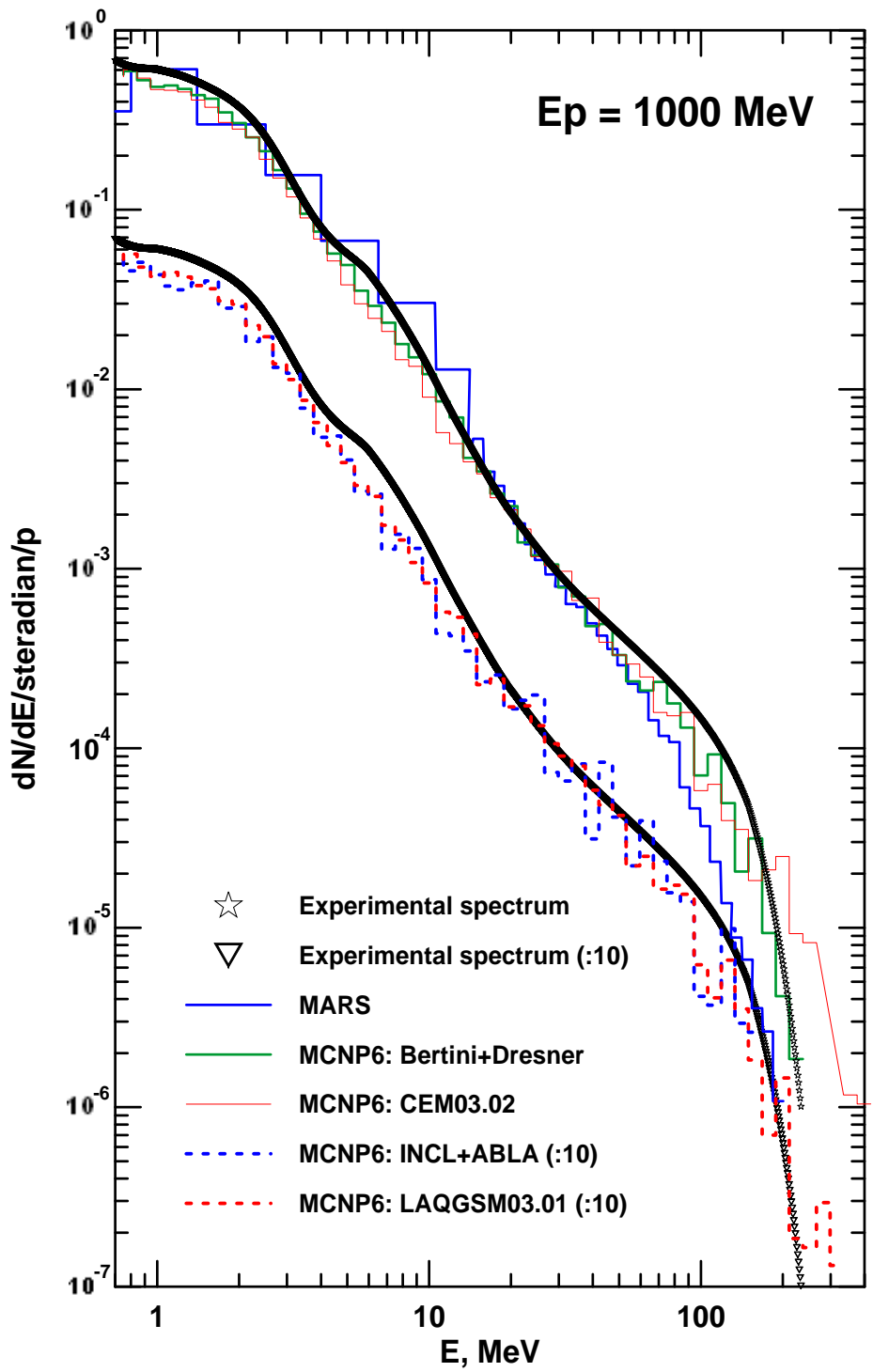

**Fig. 12.** Experimental neutron spectrum for 1000 MeV initial proton energy and calculations by the MARS and the MCNP6 codes.